\documentclass[aps,twocolumn,superscriptaddress,showpacs,10pt]{revtex4}%
\usepackage{amsfonts}
\usepackage{amsmath}
\usepackage{amssymb}
\usepackage{graphicx}%
\begin{document}
\preprint{ }
\title{Double-layer shocks in a magnetized quantum plasma}
\author{A. P. Misra}
\email{apmisra@visva-bharati.ac.in}
\affiliation{Department of Physics, Ume{\aa } University, SE--901 87 Ume{\aa }, Sweden}
\affiliation{Department of Mathematics, Siksha Bhavana, Visva-Bharati University,
Santiniketan-731 235, India.}
\author{S. Samanta }
\affiliation{Department of Basic Science and Humanities, College of Engineering and
Management, Kolaghat-721 171, India.}


\begin{abstract}
The formation of small but finite amplitude electrostatic shocks in the
propagation of quantum ion-acoustic waves (QIAWs) obliquely to an external
magnetic field is reported in a quantum electron-positron-ion (e-p-i) plasma.
Such shocks are seen to have double-layer (DL) structures \ composed of the
compressive and accompanying rarefactive slow-wave fronts. Existence of such
DL shocks depends critically on the quantum coupling parameter $H$ associated
with the Bohm potential and the positron to electron density ratio $\delta$.
The profiles may, however, steepen initially and reach a steady state with a
number of solitary waves in front of the shocks. Such novel DL shocks could be
a good candidate for particle acceleration in intense laser-solid density
plasma interaction experiments as well as in compact astrophysical objects,
e.g., magnetized white dwarfs.

\end{abstract}
\received{19 April, 2010}

\revised{24 June, 2010}

\accepted{18 August, 2010}

\startpage{1}
\endpage{102}
\maketitle

Recent studies \cite{Moslem,Khan,Misra,Chatterjee} have shown that the
formation of stationary current free double layers (DLs) may be possible in
dense plasma environments where the forces namely, (i) quantum statistical
pressures for electrons and positrons, (ii) electron and positron tunneling
associated with the Bohm potential play important roles in the propagation of
quantum ion-acoustic waves (QIAWs). Inclusion of these quantum forces along
with electron and positron angular momentum spin allows also the existence of
very high-frequency dispersive electrostatic and electromagnetic waves (e.g.,
in the hard $X$-ray and $\gamma$-ray regimes) with extremely short wavelengths
(see for recent review in quantum plasmas, Ref. \cite{Shukla}). However, there
was no indication or explanation of the forward propagating (upstream) DLs
\ that may exist, especially in a strongly magnetized quantum plasma, as a
form of compressive and accompanying rarefactive slow-wave fronts. Formation
of these upstream DLs is one of the most striking features of the
magnetohydrodynamic (MHD) shocks, and has been observed by Tajima et al
\cite{Tajima}, possibly for the first time, in a classical MHD flow. Moreover,
\ there is also a wide-spread interest in investigating DLs as a possible
acceleration mechanism in various space and astrophysical plasma environments
(see, e.g., \cite{Samsonov,Bletzinger,Bostrom,Hairapetian,Quon}). In a
thought-provoking series of discussions, Alfv\'{e}n considered DLs to be a
central paradigm in plasma astrophysics \cite{Alfven}. \ 

Our current knowledge of DLs in plasma physics, however, is insufficient for
us to judge with much confidence and belief what roles DLs may, indeed, play
especially in astrophysical environments. Thanks for the evidence of particle
acceleration in a magnetized white dwarf which has been reviewed by Jager in
the past \cite{Jager}.\ It has been reported there that if the current density
is large enough in a tenuous magnetosphere, DLs can be formed leading to a
large electric field, and hence monoenergetic electrons and ions with higher
energies. Recently, a new study (see, e.g., Ref. \cite{Whitedwarf})
demonstrates that a white dwarf star may pulse like a pulsar, e.g., AE Aquarii
can emit pulses of high energy $X$-rays as it rotates on its own axis. \ It
has also been predicted that since pulsars are known to be sources of cosmic
rays, white dwarfs should be quiet but numerous particle accelerators
\cite{particleaccelerator}, contributing many of the low-energy cosmic rays in
our galaxy. One may thus be interested to know the physical mechanisms of such
particle accelerations in these astrophysical compact objects. In this regard
it may be noted that ever since the discovery of cosmic rays, the problem of
understanding their origin and acceleration has not yet been fully understood.
In this context, a number of models and several processes have been proposed
in the early days for their origin and the acceleration mechanism (see for
some discussions, e.g., Ref. \cite{CosmicRays}). Furthermore, in addition to
the shock-wave acceleration mechanism proposed early by Colgate and white
\cite{Colgate}, Goldreich et al \cite{Goldreich} and Gunn et al \cite{Gunn}
had also shown that pulsars can accelerate particle to very high energies.
Recently, the expected abundance of cosmic-ray electrons and positrons from
pulsars and magnetars have been studied \cite{RecentCosmicRays}.

A basic prerequisite for particle acceleration in pulsars is the presence of
magnetospeheric plasma. On the basis of which Goldreich and Julien
\cite{Goldreich} had shown that the large electric field due to a nonzero
$E.B$ at the neutron star surface, can overcome the gravitational potential
and ensure a minimum plasma density. They also showed that the plasma flowing
along the co-rotating magnetic field lines and escaping out at the light
cylinder (defined by the distance from the center of the neutron star) will
experience a potential drop and hence be accelerated. Furthermore, since an
oblique magnetic rotator (in which the magnetic and rotational axes do not
coincide) emits electromagnetic (EM) waves of the same frequency as that of
rotation, particles can be accelerated at or beyond the light cylinder to high
energies extremely efficiently by the low-frequency EM waves emitted by a
pulsar. This was first proposed by Gunn et al \cite{Gunn}.

Note, however, that in addition to degenerate electrons, there would also
exist degenerate positrons, e.g., in magnetars and in the next generation of
intense laser-solid density plasma interaction experiments \cite{Shukla} for
which DLs could well be responsible for ion acceleration therein.

Motivated by these facts, we report in this work, the existence of forward
propagating slow-wave shocks in a magnetized quantum plasma, whose
constituents are electrons, positrons and positive ions (hereafter referred as
e-p-i). Because of the sufficient lifetime of the positrons compared to the
ion time scale, plasma can become an admixture of electrons, positrons and
ions. Such e-p-i plasmas are believed to exist, e.g., in the magnetosphere of
pulsars, in the active galactic nuclei, in the regions of the accretion disks
surrounding the central black holes \cite{Goldreich,Sturrock,Michel,Miller},
Van Allen radiation belts, near the polar cap of fast rotating neutron stars
\cite{Lightman}, supernova remnants \cite{Piran}, in intense laser fields
\cite{Berezhiani}, in compact astrophysical objects (e.g., giant planetary
interiors, white dwarfs, neutron stars/ magnetars) \cite{superdense}, in
tokamaks \cite{Helander} as well as in the early universe \cite{Rees}. Note
that the process of electron-positron (e-p) pair creation and annihilation may
occur in relativistic plasmas at high temperatures, when the temperature of
the plasma exceeds the rest mass of electrons. \ However, for the propagation
of ion-acoustic waves in e-p-i plasmas, the e-p pair annihilation can be
neglected in the sense that the electron-positron lifetime is much larger than
the characteristic time scale for collective oscillations (see for detail
derivation and discussion, e.g., Ref. \cite{Annihilation}).

In what follows, we will consider the quantum force associated with the Bohm
potential to provide higher order dispersion along with the charge separation
effect as well as the magnetic-field-induced dispersion anisotropy. The ions
are assumed to be cold and the motion is considered on the ion-acoustic time
scale. Because of their light masses, electrons and positrons will be highly
magnetized compared to the ions ( ion Larmor radius is much larger than that
of electrons or positrons), and will move almost parallel to the external
magnetic field, so that electrons and positrons may be described by the
quantum modified Boltzmann-like distributions. We will derive a modified
Korteweg-de Vries (MKdV) equation with a quadratic as well as cubic
nonlinearity that describes the dynamics of quantum ion-acoustic (QIA) DLs,
and investigate some interesting properties of such DLs for large times. We
observe a novel DL structure to the upstream shocks, which is composed of a
compressive slow-wave shock and a rarefactive slow-wave front.

Under the above assumptions, the motions of ions, electrons and positrons in
the propagation of QIAWs obliquely to the external magnetic field $B=B_{0}%
\hat{z}$ can be described by the following equations \cite{Haas}.
\begin{equation}
\partial_{t}n_{i}+\nabla.(n_{i}\mathbf{v})=0, \label{1}%
\end{equation}%
\begin{equation}
\left(  \partial_{t}+\mathbf{v}.\nabla\right)  \mathbf{v}=-\nabla\phi
+\omega_{c}\mathbf{v}\times\hat{z}, \label{2}%
\end{equation}%
\begin{equation}
n_{e}^{2/3}=1+2\phi+\frac{H^{2}}{\sqrt{n_{e}}}\nabla^{2}\sqrt{n_{e}},
\label{3}%
\end{equation}

\begin{equation}
n_{p}^{2/3}=1-2\delta^{-2/3}\phi+\frac{H^{2}}{\sqrt{n_{p}}}\nabla^{2}%
\sqrt{n_{p}}, \label{eq3}%
\end{equation}

\begin{equation}
\nabla^{2}\phi=n_{e}-\delta n_{p}-\left(  1-\delta\right)  n_{i}, \label{4}%
\end{equation}
where $n_{\alpha}$ is the number density of $\alpha-$species particle
normalized by the unperturbed value $n_{\alpha0},$ $v\equiv\left(  v_{x}%
,v_{y},v_{z}\right)  $ is the ion fluid velocity normalized by the
ion-acoustic speed $c_{s}=\sqrt{k_{B}T_{Fe}/m_{i}}$ with $k_{B}$ denoting the
Boltzmann constant, $T_{Fe}$ the electron Fermi temperature and $m_{i}$ the
ion mass. Also, $\phi$ is the electrostatic potential normalized by
$k_{B}T_{Fe}/e,$ $\omega_{c}=eB_{0}/m_{i}\omega_{pi}$ is the ion-cyclotron
frequency normalized by the ion plasma frequency, $\omega_{pi}=\sqrt
{n_{i0}e^{2}/\epsilon_{0}m_{i}},$ $\delta=n_{p0}/n_{e0}$ is the positron to
electron density ratio, $H=\hbar\omega_{pe}/k_{B}T_{Fe}$ is the quantum
parameter denoting the ratio of the electron plasmon energy density to the
Fermi thermal energy. The space and time variables are normalized by the Fermi
Debye length $\lambda_{F}=c_{s}/\omega_{pi}$ and the ion plasma period
$\omega_{pi}^{-1}$ respectively. In Eqs. (\ref{3}), (\ref{eq3}) we have used
the following Fermi-Dirac pressure law due to electron and positron degeneracy
\cite{Landau}.
\begin{equation}
P_{\alpha}=\frac{1}{5}\frac{m_{\alpha}V_{F\alpha}^{2}}{n_{\alpha0}^{2/3}%
}n_{\alpha}^{5/3}, \label{5}%
\end{equation}
where $V_{F\alpha}=\sqrt{k_{B}T_{F\alpha}/m_{\alpha}}$ is the Fermi thermal
speed. In Eqs. (\ref{3}), (\ref{eq3}), the terms proportional to $H$
associated with the Bohm potential, account for typical quantum effects such
as tunneling. In a broad sense, we refer to these particularities arising from
the wave-like nature of the charge carriers as `quantum diffraction effects'.

In order to describe now the dynamics of propagating DLs, we employ a
reductive perturbation technique in which the independent variables are
stretched \ as $\xi=\epsilon(l_{x}x+l_{y}y+l_{z}z-Mt),$ $\tau=\epsilon^{3}t,$
where $\epsilon$ is a small parameter representing the strength of the wave
amplitude, $M$ is the phase speed normalized by $c_{s}$ and $l_{x},l_{y}%
,l_{z}$ are the direction cosines of the wave vector along the axes such that
$l_{x}^{2}+l_{y}^{2}+l_{z}^{2}=1$. The dependent variables are expanded as
$n_{\alpha}=1+\sum\limits_{j=1}^{\infty}\epsilon^{j}n_{\alpha j},$ $\left(
v_{x},v_{y}\right)  =\sum\limits_{j=1}^{\infty}\epsilon^{1+j/2}\left(
v_{xj},v_{yj}\right)  ,\left(  v_{z},\phi\right)  =\sum\limits_{j=1}^{\infty
}\epsilon^{j}\left(  v_{zj},\phi_{j}\right)  $. Here the transverse velocity
components $\left(  v_{x},v_{y}\right)  $ appear at higher order of $\epsilon$
than that of the parallel component $v_{z}$. This anisotropy is introduced by
the influence of strong magnetic field. In this approximation the ion
gyromotion is treated as higher order effect. By inserting these expressions
into Eqs. (\ref{1})-(\ref{4}) and collecting the terms in different powers of
$\epsilon$, we obtain in the lowest order, $n_{i1}=\left(  l_{z}/M\right)
^{2}\phi_{1},$ $n_{e1}=3\phi_{1},$ $n_{p1}=-3\delta^{-2/3}\phi_{1},$
$v_{z1}=\left(  l_{z}/M\right)  \phi_{1},$ $v_{x1}=v_{y1}=0$ \ together with
the dispersion law%

\begin{equation}
M=\pm l_{z}\sqrt{\frac{(1-\delta)}{3(1+\alpha)}},
\end{equation}
where $\alpha\equiv\delta^{1/3}.$ \ \ The flow may be outward or inward
depending on the sign we consider in Eq. (\ref{4}). \ Moreover, $M<1,$ i.e.,
the QIAWs propagate with the phase speed smaller than the ion-acoustic speed,
and $M$ increases as the positron to electron density ratio increases.

Next, the coefficient of $\epsilon^{2}$ vanishes leading to $Q\phi_{1}^{2}=0$
where $Q=1/\alpha-9(1+\alpha)/\left(  1-\delta\right)  .$ Without loss of
generality we may assume that $\epsilon\phi_{1}\rightarrow-\phi_{0}/2$ as
$r\equiv(x,y,z)\rightarrow\infty$ so that $\phi_{2}\rightarrow0$ as
$r\rightarrow\infty.$ Thus, $Q$ should be at least of the order of $\epsilon,$
i.e, $Q\phi_{1}^{2}\sim\epsilon^{3}$ and is to be added to the third order
contribution from Eq. (\ref{4}). This gives a favorable condition for DL
shocks instead of solitons. Physically, under this condition the free as well
as the trapped particles are assumed to adjust themselves rapidly in order to
maintain the quasineutrality\ at any time on each side of the propagating
shocks. Now, the vanishing of the coefficients of $\epsilon^{3}$ gives five
equations. When $\partial\phi_{3}/\partial\xi,$ $\partial n_{i3}/\partial\xi,$
$\partial v_{z3}/\partial\xi$ etc. are eliminated and the first-order
quantities are inserted into the resulting equation, terms containing
$\phi_{2}$ and $\phi_{3}$ cancel, and the following MKdV equation is obtained
$\left(  \phi\cong\epsilon\phi_{1}\right)  $.%
\begin{equation}
\frac{\partial^{3}\phi}{\partial\xi^{3}}=\lambda\frac{\partial\phi}%
{\partial\tau}+2\mu\phi\frac{\partial\phi}{\partial\xi}+3\gamma\phi^{2}%
\frac{\partial\phi}{\partial\xi}, \label{7}%
\end{equation}
where $\lambda=P/S,$ $\mu=Q/S,$ $\gamma=R/S$ with $P=6(1+\alpha)/l_{z},$
$R=135(1+\alpha)^{3}/\left(  1-\delta\right)  ^{2}-3\left(  1-\delta\right)
/2\delta$ and $S=1+(1-\delta)(1-l_{z}^{2})/\omega_{c}^{2}-9H^{2}%
(1+\alpha)/4\alpha.$ An asymptotic shock solution of Eq. (\ref{7}) can be
obtained as \cite{Torven}
\begin{equation}
\phi\left(  \xi,\tau\right)  =-\frac{\phi_{0}}{2}{\tanh}\left(  -Ns_{1}%
+\sqrt{-\frac{\gamma}{8}}\phi_{0}\zeta\right)  \label{10}%
\end{equation}
where $\zeta=\xi-V\tau,$ $s_{1}$ is a constant and $N-1$ represents the number
of solitary waves in front of the shocks. Let us first investigate
analytically the coefficients $\lambda,$ $\mu$ and $\gamma$ with the system
parameters. We see that for $\delta<1,$ $P\ $\ is always negative and
$Q(\sim\epsilon)\gtrless0$ according as $0<\delta<0.001$ or $0.001<\delta<1.$
Also, $R$ is negative for $0<\delta<0.0065$ and positive otherwise. Moreover,
$S\gtrless0$ according as $H\gtrless H_{c},$ where $H_{c}$ is the critical
value of $H$ given by%

\begin{equation}
H_{c}=\frac{2}{3}\sqrt{\frac{\alpha}{1+\alpha}\left[  1+\frac{(1-\delta
)(1-l_{z}^{2}}{\omega_{c}^{2}}\right]  }. \label{11}%
\end{equation}
Note that this critical value, which depends parametrically on the density
ratio $\delta,$ the obliqueness parameter $l_{z}$ and the ion-cyclotron
frequency $\omega_{c}$ gives a critical value of the electron density. The
smaller the values of $H_{c}\ $the larger are the electron number densities.
In order to consider the smaller values of $H_{c}$ or higher densities, one
might have to disregard the charge separation effect (the unity in the square
brackets) in strongly magnetized plasmas $(\omega_{c}<1)$. In this case the
critical value scales as \ $H_{c}\sim2\sqrt{\alpha/\left(  1+\alpha\right)
}/3.$ In a weakly magnetized case, $H_{c}\sim2\sqrt{\alpha(1-\delta
)(1-l_{z}^{2})/\left(  1+\alpha\right)  }/3\omega_{c},$ which can relatively
be larger than the strongly magnetized case. However, in both the cases one
has to be careful about the particle density range in which \ the Fermi
thermal speed is much smaller than the speed of light in vacuum ($V_{F\alpha
}<<c$) and the coupling parameter satisfies the relation: $g_{Q}\equiv
2m_{e}e^{2}/\varepsilon_{0}\hbar^{2}(3\pi^{2}\sqrt{n_{0}})^{2/3}\lesssim1$
(this corresponds to the case where the quantum collective and mean field
effects are important). Since $Q\sim\epsilon$ for $\delta\sim0.001,$ we
consider the regime $0<\delta<0.0065$ in which $R<0.$ Thus, in order that the
DL solutions exist we must have $H>H_{c}$ such that $S>0,$ since $\gamma$ is
to be negative.

Note that since $S$ is always positive for $\delta,$ $l_{z}$ $<1$ and for
$H=0$, the DL solutions still exist in the absence of $H$. This implies that
quantum effects will be relevant in dense plasma environments where electrons
and positrons are degenerate (e.g., in magnetars as well in the next
generation laser solid-density plasma interaction experiments) for which
classical fluid model fails to describe the plasma dynamics. Basically, the
quantum parameter $H$ $\left(  >H_{c}\right)  $ \ restricts here the particle
density to be of the order of $10^{34}$m$^{-3}\ $or lower (since higher values
of $H$ corresponds to lower density regimes) in order that the ion-acoustic
DLs exist. Furthermore, the DLs are compressive or rarefactive according as
$\delta\lessgtr0.001.$ Inspecting Eq. (\ref{10}) one finds that while the
width decreases, the amplitude of the stationary DLs increases with a slight
increase of the density ratio $\delta.$ Also, the width increases, with the
frequency $\omega_{c}$ and the obliqueness $l_{z}.$

Next, we numerically investigate Eq. (\ref{7}) using Runge-Kutta scheme with
an initial condition $\phi\left(  \xi,0\right)  =-\left(  \phi_{0}/2\right)
\tanh\left[  \sqrt{-\gamma/8}\left(  \phi_{0}/N\right)  \xi\right]  $ and with
$\phi_{0}\sim-0.05,$ $N=3,$ $\delta\sim0.001,$ $H\sim0.22.$ The numerical
values of the coefficients in Eq. (\ref{7}) are $\lambda=-25.5,$ $\mu
=0.17\ $and $\gamma=-2.6\times10^{3}.$ Note here that though the initial
condition looks very similar to the asymptotic solution (\ref{10}), but the
constant $N$ appears in different manner (For detailed discussion see, e.g.,
Ref. \cite{Torven}). We use $1000$ grid points with the system scale size
$L_{\xi}=100,$ and choose the pulse size $L_{p}(\sim1.2)$ to be less than
$L_{\xi}$ in order that the shock solutions exist. The shock profiles are
shown in Fig. 1 giving a spatial scaling constant $(3.7\lambda_{F})^{-1}$ for
the initial profile and $(1.2\lambda_{F})^{-1}$ for the asymptotic shock. The
solitary waves in front of the shocks are ordered so that one with the maximum
amplitude is nearest the shock. Initially the shock profile steepens, as time
\ goes on it reaches a steady state with two solitary wave fronts (in the case
of $N=3$). \begin{figure}[ptb]
\begin{center}
\includegraphics[height=2.8in,width=3.5in]{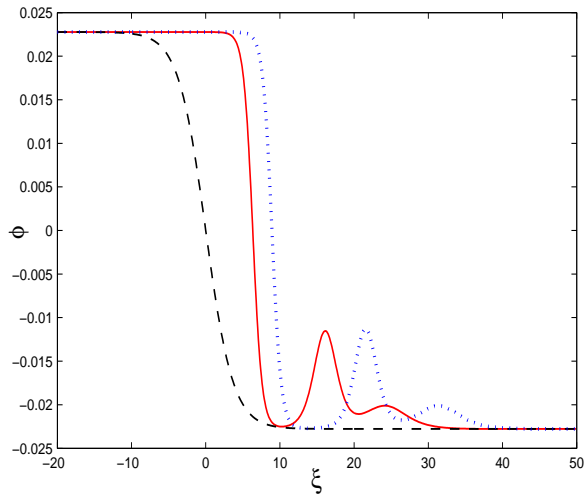}
\end{center}
\caption{(Color online) The profiles of the DL shocks for three different
times. The initial profile ($\tau=0$) steepens (dashed line), and
simultaneously two solitary waves form in front of the shock (see the solid
line for $\tau=150$ and the dotted line for $\tau=200$).}%
\end{figure}

In conclusion, the formation of forward propagating DL shocks is possible in a
strongly magnetized quantum e-p-i plasma. Such shocks composed of the
compressive as well as rarefactive slow-wave fronts propagate with the speed
less than that of the ion-acoustic speed. Furthermore, the DLs exist in dense
plasma environments with particle density$\ $of the order of $10^{34}$%
m$^{-3}\ $or lower when the background electron population is much larger than
that of positrons, and electron-positron annihilation is negligible
\cite{Annihilation}. Existence of such DL shocks may have significant role for
the particle acceleration in the next generation laser solid-density plasma
interaction experiments \cite{Shukla} as well as in compact astrophysical
objects as evident from the recent observations in magnetized white dwarfs
\cite{Whitedwarf}. However, conclusive evidence needs further investigation in
this area. In this way one may extend our investigation by considering the
relativistic as \ well as the spin quantum effects \cite{spin} in a quantum
MHD model, which, we hope, will give better understanding for the existence
and properties of such novel DL structures. Future research is also expected
to reveal other interesting applications of the shock solutions demonstrated here.

S. S. is thankful to Professor A. Roy Chowdhury of Department of Physics,
Jadavpur University, Kolkata-700 032, India, for some useful discussions. A.
P. M. is grateful to the Kempe Foundations, Sweden, for support.


\begin{thebibliography}{99}
\bibitem {particleaccelerator}The white dwarf acts like a particle accelerator
when its magnetic field lines break up and accelerate particles to
relativistic speeds. Electrons then spiral out along these broken magnetic
field lines.
\end{thebibliography}
\end{document}